\documentclass[twocolumn, aps, pra, 10pt]{revtex4-2}

\usepackage{graphicx}

\usepackage{amsmath, amssymb, amsfonts}

\usepackage{hyperref}
\usepackage{color}
\usepackage{xcolor}
\begin{document}


    \title{Mechanical Squeezing in Quadratically-coupled Optomechanical Systems}
    \author{Priyankar Banerjee}
        \affiliation{Department of Physics, Indian Institute of Technology Guwahati, Guwahati-781039, India}
    \author{Sampreet Kalita}
        \affiliation{Department of Physics, Indian Institute of Technology Guwahati, Guwahati-781039, India}
    \author{Amarendra K. Sarma}
        \email{aksarma@iitg.ac.in}
        \affiliation{Department of Physics, Indian Institute of Technology Guwahati, Guwahati-781039, India}

    \begin{abstract}
        We demonstrate the generation of a strong mechanical squeezing in a dissipative optomechanical system by introducing a periodic modulation in the amplitude of a single-tone laser driving the system. 
        The mechanical oscillator is quadratically coupled to the optical mode, which contributes to a strong squeezing exceeding the 3-dB standard quantum limit.
        The Bogoliubov mode of the mechanical oscillator also cools down to its ground state due to sideband cooling.
        We further optimize this ratio of sideband strengths to introduce enhanced squeezing.
        We also compare our results with the analytical (under adiabatic approximation) and the exact numerical solution.
        Even for a thermal occupancy of $10^{4}$ phonons, mechanical squeezing beyond 3 dB and a strong optomechanical entanglement is observed.
    \end{abstract}

    \maketitle


    \section{Introduction}
        \label{sec:intro}        
        In recent years, the amount of research on macroscopic quantum systems for the generation of nonclassical states such as squeezed and entangled states has increased manifold \cite{PhysRevA.101.053836, PhysRevA.104.053506, PhysRevA.93.043844, PhysRevA.99.043805, PhysRevLett.117.100801, PhysRevLett.115.243601}.
        Such states have numerous applications in the development of quantum technologies for quantum computation \cite{PhysRevLett.90.137901}, quantum information processing \cite{RevModPhys.77.513, PhysRevA.84.052327}, and quantum-enhanced force sensing \cite{Science.304.74, NewJPhys.18.073040}.
        In the past decades, micro-mechanical systems relying on position measurements have been extensively used to sense, store and process information.
        However, the precision of position-measurements of a mechanical oscillator is limited by its zero-point fluctuations \cite{TaylorFrancis.QuantumOptomechanics.Bowen}. 
        In addition to that, measuring the mechanical position may itself induce noise into the system.
        In fact, most position measuring systems have noise contributions that exceed the standard quantum limit (SQL) of $3$ dB \cite{Science.349.952}.
        To circumvent this, the motion of the mechanical oscillator can be squeezed beyond the SQL, by reducing the variance in one of its quadratures at the expense of increased variance in the other quadrature \cite{PhysRevA.46.R1181}.
        Also, the preparation and control of a mechanical element in a quantum state of motion at a mesoscopic level allows us to test quantum mechanics' fundamental hypotheses at the quantum-classical boundary \cite{SBH.CavityOptomechanics.Aspelmeyer}.
        A suitable platform to generate such quantum states is a cavity optomechanical system \cite{RevModPhys.86.1391}.
        Such a system relies on the radiation-pressure interaction of a mechanical system with the electromagnetic radiation inside a cavity.
        The high sensitivity of the phase of the cavity photons to small displacements of the mechanical oscillator makes them an ideal candidate for force-sensing applications and are routinely used in interferometric detectors such as the ones in LIGO \cite{RepProgPhys.72.076901}, and VIRGO \cite{AstropartPhys.34.521} experiments.
        
        Recently, many optomechanical models have been proposed to induce squeezing below the 3 dB level by using squeezed light \cite{PhysRevA.82.033811, PhysRevA.79.063819}, two-tone driving \cite{Science.349.952, PhysRevA.88.063833, PhysRevA.99.043805}, reservoir engineering \cite{PhysRevA.88.013835, PhysRevA.79.052102, PhysRevLett.103.213603, PhysRevB.70.205304} and frequency modulation \cite{PhysRevLett.117.100801}.
        It has also been shown that periodically modulating the amplitude of the external field can induce a high degree of mechanical squeezing and optomechanical entanglement in a cavity optomechanical system \cite{PhysRevLett.103.213603}.
        A prerequisite of such a scheme is the ground-state cooling of the mechanical mode, usually achieved by sideband cooling.
        Dissipative optmechanical systems, where the mechanical mode modulates the decay rate of the cavity mode \cite{PhysRevLett.102.207209} enables this in the unresolved sideband limit \cite{PhysRevLett.102.207209, PhysRevLett.107.213604, PhysRevLett.103.223901, PhysRevA.88.023850, PhysRevA.102.043520}.
        Such a scheme can be thought as an example of reservoir engineering \cite{PhysRevLett.77.4728} where the cavity acts as a reservoir whose force noise is squeezed.
        
        Most schemes implemented to achieve mechanical squeezing using reservoir engineering technique are based on two-tone driving \cite{PhysRevLett.115.243601, PhysRevA.88.063833, PhysRevLett.117.100801, Science.349.952}.
        Only recently, Bai et. al. \cite{PhysRevA.101.053836} proposed a scheme to squeeze the mechanical position using single-tone driving in a standard optomechanical system.
        In this work, we study the dissipative generation of mechanical squeezing in a quadratically-coupled optomechanical system by periodically modulating the driving amplitude and show that along with a robust squeezing, a high degree of optomechanical entanglement can be produced.
        We also show that the steady-state squeezing generated using this scheme is sensitive to the ratio of the cavity drive amplitudes.
        The paper is organised as follows.
        In Sec. \ref{sec:system}, we introduce the membrane-in-the-middle cavity optomechanical system driven by a amplitude-modulated single-tone laser, analyzing the dynamical behavior of the system with and without the rotating-wave approximation (RWA). 
        In Sec. \ref{sec:squeeze}, we present the squeezing in the position quadrature of the mechanical mode and discuss the squeezing effect as a competing behaviour between two conflicting tendencies, and obtain an optimal ratio of the sidebands to maximize squeezing by balancing these effects.
        In Sec. \ref{sec:soln}, we derive an analytical solution for squeezing of the mechanical position under adiabatic approximation, and compare it with the exact numerical solution.
        We then examine the robustness of the squeezing achieved and the behaviour of entanglement between the optical and mechanical modes in Sec. \ref{sec:robust}.
        Finally, we summarise our paper in Sec. \ref{sec:conc}.


    \section{System and Dynamics}
        \label{sec:system}
        \begin{figure}[ht]
            \centering \includegraphics[width=0.48\textwidth]{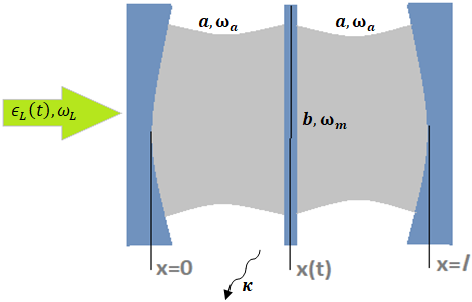}
            \caption{A membrane-in-the-middle optomechanical system, driven by an amplitude-modulated external laser.}
            \label{fig:system}
        \end{figure}

        The system under consideration is illustrated in Fig. \ref{fig:system}.
        An optomechanical system is driven by an external laser (frequency $\omega_{l}$) whose amplitude $\epsilon_{l}$ is periodically modulated.
        The optical mode (frequency $\omega_{o}$) is coupled to the mechanical mode (frequency $\omega_{m}$) via a radiation-pressure interaction of strength $g$.
        In a frame rotating at the frequency $\omega_{l}$, the system Hamiltonian takes the form (in units of $\hbar$) \cite{RevModPhys.86.1391}
        \begin{eqnarray}
            \label{eqn:system_ham}
            H & = & \Delta_{0} a^{\dagger} a + \omega_{m} b^{\dagger} b - g a^{\dagger} a \left( b^{\dagger} + b \right)^{2} \nonumber \\
            && + i \epsilon_{l} \left( a^{\dagger} - a \right) + \eta \omega_{m} \left( b^{\dagger} + b \right),
        \end{eqnarray}
        where $\Delta_{0} = \omega_{o} - \omega_{l}$ is the detuning of the optical cavity and $a$ ($b$) is the annihilation operator of the optical (mechanical) mode.
        Here, the first and the second terms represent the individual energies of the optical cavity and the mechanical membrane.
        The third term is the optomechanical interaction energy with quadratic coupling.
        The fourth term represents the time-dependent laser whose driving amplitude is such that $\epsilon_{l} (t) = \epsilon_{l} (t + \tau) = \sum_{n = - \infty}^{n = \infty} \epsilon_{n} e^{-i n \Omega t}$, where $\tau$ is the modulation period with $\Omega = 2 \pi / \tau$, and $\epsilon_{n}$'s are the sideband modulation strengths.The final term represents a constant impulsive force of $\eta \hbar \omega_{m}$ acting on the mechanical membrane.

        The time evolution of the mode operators of the system follow the Heisenberg equations of motion \cite{TaylorFrancis.QuantumOptomechanics.Bowen}.
        Taking into account the dissipation of the cavity mode (rate $\kappa$) and the decay of the mechanical resonator (rate $\gamma$), along with the effect of the vacuum and thermal fluctuations entering the system, we obtain the quantum Langevin equations (QLEs) given by \cite{PhysRevA.31.3761}
        \begin{subequations}
            \label{eqn:system_qle}
            \begin{eqnarray}
                \dot{a} & = & - i \Delta_{0} a - \frac{\kappa}{2} a + i g a \left( b^{\dagger} + b \right)^{2} + \epsilon_{l}+ \sqrt{\kappa} a_{in}, \\
                \dot{b} & = & - i \omega_{m} b - \frac{\gamma}{2} b + 2 i g a^{\dagger} a \left( b^{\dagger} + b \right) - i \eta \omega_{m} \nonumber \\
                && + \sqrt{\gamma} b_{in}.
            \end{eqnarray}
        \end{subequations}

        Here, $a_{in}$ and $b_{in}$ are the noise operators associated with the vacuum and thermal fluctuations, which are assumed to have a Gaussian nature, given as \cite{SBH.CavityOptomechanics.Aspelmeyer, SBH.QuantumOptics.Walls}
        \begin{subequations}
            \label{correlations}
            \begin{eqnarray}
                \langle a_{in}^{\dagger}(t)a_{in}(t')\rangle & = & n_{a}\delta(t-t'), \\
                \langle a_{in}(t)a_{in}^{\dagger}(t')\rangle & = & (n_{a}+1)\delta(t-t'), \\
                \langle b_{in}^{\dagger}(t)b_{in}(t')\rangle & = & n_{b}\delta(t-t'), \\
                \langle b_{in}(t)b_{in}^{\dagger}(t')\rangle & = & (n_{b}+1)\delta(t-t'),
            \end{eqnarray}
        \end{subequations}
        where $n_{a}$ and $n_{b}$ are the thermal occupancies of the optical and mechanical modes respectively, given by $n_{j} = \{ \textrm{exp} [ \hbar \omega_{j} / ( k_{B} T ) ] - 1 \}^{-1}$ for $j \in \{ a, b \}$ at temperature $T$.
       
        For sufficiently strong drive amplitudes, the QLEs obtained in Eqs. \eqref{eqn:system_qle} can be approximated using a linearized description by assuming that the mode operators ($\mathcal{O}$) are equal to the sum of their classical expectation values and their quantum fluctuations, i.e., $ \mathcal{O} = \langle \mathcal{O} \rangle + \delta \mathcal{O}$ \cite{TaylorFrancis.QuantumOptomechanics.Bowen}.
        This gives us the classical equations for $\alpha = \langle a \rangle$ and $\beta = \langle b \rangle$ as
        \begin{subequations}
            \label{eqn:system_class}
            \begin{eqnarray}
                \dot{\alpha} & = & - i \Delta_{0} \alpha - \frac{\kappa}{2} \alpha + i g \alpha \left( \beta^{*} + \beta \right)^{2} + \epsilon_{l}, \\
                \dot{\beta} & = & - i \omega_{m} \beta -\frac{\gamma}{2} \beta + 2 i g \left| \alpha \right|^{2} \left( \beta^{*} + \beta \right) - i \eta \omega_{m},
            \end{eqnarray}
        \end{subequations}
        and their quantum counterparts ($\delta \mathcal{O} \rightarrow \mathcal{O}$)
        \begin{subequations}
            \label{eqn:system_quant}
            \begin{eqnarray}
                \dot{a} & \approx & - i \Delta a - \frac{\kappa}{2} a + 2 i g \alpha \left( \beta^{*} + \beta \right) \left( b^{\dagger} + b \right)  + \sqrt{\kappa} a_{in}, \\
                \dot{b} & \approx & - i \omega_{m} b -\frac{\gamma}{2} b + 2 i g \left( \alpha^{*} a + \alpha a^{\dagger} \right) \left( \beta^{*} + \beta \right) \nonumber \\ 
                && + 2 i g \left| \alpha \right|^{2} \left( b^{\dagger} + b \right) + \sqrt{\gamma} b_{in},
            \end{eqnarray}
        \end{subequations}
        where, the effective detuning of the linearized dynamics is given by $\Delta = \Delta_{0} - g ( \beta^{*} + \beta )^{2}$.
        
        As the primary contribution of the modulated drive comes from the offset strength and the first-order modulations, for our analysis, we assume that $\epsilon_{l} (t) \approx \epsilon_{-1} e^{i \Omega t} + \epsilon_{0} + \epsilon_{1} e^{- i \Omega t}$.
        Then, according to the Floquet theorem, at a long-time limit, the cavity mode and the mechanical mode amplitude would show the same behaviour as the modulated external field, i.e., $\lim_{t \to \infty} \alpha (t) = \alpha (t + \tau)$ and $\lim_{t\to\infty} \beta (t) = \beta (t + \tau)$ \cite{PhysRevLett.103.213603, PhysRevA.101.053836, PhysRevA.104.053506}. 
        We therefore redefine these classical amplitudes as
        \begin{subequations}
            \label{eqn:system_amp}
            \begin{eqnarray}
                \alpha & = & a_{-1} e^{i \Omega_{a} t} + a_{0} + a_{1} e^{- i \Omega_{a} t}, \\
                \beta  & = & b_{-1} e^{i \Omega_{b} t} + b_{0} + b_{1} e^{- i \Omega_{b} t}.
            \end{eqnarray}
        \end{subequations}

        \subsection{Dynamics under RWA}
            \label{subsec:dyna_rwa}
            We now analyze the dynamics of the slowly varying fluctuations in the rotating frame of their oscillations.
            We rewrite the fluctuation operators and their corresponding input noises as $a = \tilde{a} e^{-i \Delta t}$, $b = \tilde{b} e^{-i \omega_{m} t}$ and $a_{in} = \tilde{a}_{in} e^{-i \Delta t}$, $b_{in} = \tilde{b}_{in} e^{-i \omega_{m} t}$ respectively.
            Setting the effective cavity detuning at $\omega_{m}$ and the external driving frequencies at $2 \omega_{m}$, for a weak optomechanical coupling strength, we obtain the linearized QLEs as
            \begin{subequations}
                \label{eqn:system_rwa_qles}
                \begin{eqnarray}
                    \label{eqn:system_rwa_qle_a} 
                    \dot{\tilde{a}} & = & i G_{0} \tilde{b} + i G_{1} \tilde{b}^{\dagger} - \frac{\kappa}{2} \tilde{a} + \sqrt{\kappa} \tilde{a}_{in}, \\
                    \label{eqn:system_rwa_qle_b} 
                    \dot{\tilde{b}} & = & i G_{0} \tilde{a} + i G_{1} \tilde{a}^{\dagger} + i \tilde{G}_{0} \tilde{b} + i \tilde{G}_{1} \tilde{b}^{\dagger} - \frac{\gamma}{2} \tilde{b} \nonumber \\ 
                    && + \sqrt{\gamma} \tilde{b}_{in},
                \end{eqnarray}
            \end{subequations}
            where,
            \begin{subequations}
                \begin{eqnarray}
                    G_{0} & = & 2 g \left\{ 2 a_{0} b_{0} + \left( a_{-1} + a_{1} \right)  \left( b_{-1} + b_{1} \right) \right\}, \\
                    G_{1} & = & 2 g \left\{ a_{0} \left( b_{-1} + b_{1} \right) + 2 a_{1} b_{0} \right\}, \\
                    \tilde{G}_{0} & = & 2 g \left( a_{0}^{2} + a_{-1}^{2} + a_{1}^{2} \right), \\
                    \tilde{G}_{1} & = & 2 g a_{0} \left( a_{-1} + a_{1} \right).
                \end{eqnarray}
            \end{subequations}

            In doing so, we have ignored the highly oscillatory terms under RWA.
            Using the quadrature fluctuation operators for the optical (mechanical) modes as $\tilde{X} = ( \tilde{a}^{\dagger} + \tilde{a} ) / \sqrt{2}$ ($\tilde{Q} = ( \tilde{b}^{\dagger} + \tilde{b} ) / \sqrt{2}$) and $\tilde{Y} = i ( \tilde{a}^{\dagger} - \tilde{a} ) / \sqrt{2}$ ($\tilde{P} = i ( \tilde{b}^{\dagger} - \tilde{b} ) / \sqrt{2}$) and their corresponding noise operators as $\tilde{X}_{in} = ( \tilde{a}^{\dagger}_{in} + \tilde{a}_{in} ) / \sqrt{2}$ ($\tilde{Q}_{in} = ( \tilde{b}^{\dagger}_{in} + \tilde{b}_{in} ) / \sqrt{2}$) and $\tilde{Y}_{in} = i ( \tilde{a}^{\dagger}_{in} - \tilde{a}_{in} ) / \sqrt{2}$ ($\tilde{P}_{in} = i ( \tilde{b}^{\dagger}_{in} - \tilde{b}_{in} ) / \sqrt{2}$), Eqs. \eqref{eqn:system_rwa_qles} can be written in a compact form as
            \begin{eqnarray}
                \label{eqn:system_rwa_u} 
                \dot{\tilde{\textbf{u}}} = \tilde{\textbf{M}} \tilde{\textbf{u}} + \tilde{\textbf{n}},
            \end{eqnarray}
            where the vector of fluctuations $\tilde{\textbf{u}} = ( \tilde{X}, \tilde{Y}, \tilde{Q}, \tilde{P} )^{T}$, their noises $\tilde{\textbf{n}} = ( \sqrt{\kappa} \tilde{X}_{in}, \sqrt{\kappa} \tilde{Y}_{in}, \sqrt{\gamma} \tilde{Q}_{in}, \sqrt{\gamma} \tilde{P}_{in} )^{T}$ and
            \begin{eqnarray}
                \label{eqn:system_rwa_M} 
                \tilde{\textbf{M}} = \begin{pmatrix}
                    -\frac{\kappa}{2} & 0 & 0 & -G_{-} \\
                    0 & -\frac{\kappa}{2} & G_{+} & 0 \\
                    0 & -G_{-} & -\frac{\gamma}{2} & -\tilde{G}_{-} \\
                    G_{+} & 0 & \tilde{G}_{+} & -\frac{\gamma}{2}
                \end{pmatrix},
            \end{eqnarray}
            with $G_{\pm} = G_{0} \pm G_{1}$ and $\tilde{G}_{\pm} = \tilde{G}_{0} \pm \tilde{G}_{1}$.
            
        \subsection{Dynamics without RWA}
            \label{subsec:system_wrwa}
            If we do not ignore the fast-rotating terms under RWA, the time evolution of quadrature fluctuations can be written as $\dot{\textbf{u}} = \textbf{M} \textbf{u} + \textbf{n}$, where $\textbf{u}$ is the vector of quadrature fluctuations for mode operators $a$ and $b$, and $\textbf{n}$ their corresponding noises.
            The drift matrix is then given by
            \begin{eqnarray}
                \label{eqn:system_wrwa_M} 
                \textbf{M} = \begin{pmatrix}
                    -\frac{\kappa}{2} & \Delta & -8 g \alpha_{I} \beta_{R} & 0 \\
                    -\Delta & -\frac{\kappa}{2} & 8 g \alpha_{R} \beta_{R} & 0 \\
                    0 & 0 & -\frac{\gamma}{2} & \omega_{m} \\
                    8 g \alpha_{R} \beta_{R} & 8 g \alpha_{I} \beta_{R} & -\omega_{m} + 4 g \left| \alpha \right|^{2} & -\frac{\gamma}{2}
                \end{pmatrix},
            \end{eqnarray}
            where $\alpha_{R}$ ($\beta_{R}$) and $\alpha_{I}$ ($\beta_{I}$) are the real and imaginary components of $\alpha$ ($\beta$) respectively.
            
            The correlations between the position and momentum quadrature fluctuations can then be written in terms of the correlation matrix $\textbf{V}_{kk'} = \langle u_{k} u_{k'} + u_{k} u_{k'} \rangle / 2$.
            These correlations obey the equation of motion $\dot{\textbf{V}} = \textbf{M} \textbf{V} + \textbf{V} \textbf{M}^{T} + \textbf{D}$, where $\textbf{D} = \mathrm{Diag} [ \kappa ( n_{a} + 1 / 2 ), \kappa ( n_{a} + 1 / 2 ), \gamma ( n_{b} + 1 / 2 ), \gamma ( n_{b} + 1 / 2 ) ]$ is the noise matrix. 
            The diagonal elements $\textbf{V}_{33}(t)$ and $\textbf{V}_{44}(t)$ represent the variance of the position and momentum quadrature of the mechanical mode, respectively.
            
            It can be noted here that the eigenvalues of $\tilde{\textbf{M}}(t)$ have to be negative and real for the system to be stable.
            To do this, we follow the Routh-Hurwitz criteria \cite{SBH.ControlTheory.Bellman} and obtain the following inequalities for system stability to hold:
            \begin{subequations}
                \label{eqn:system_rhc}
                \begin{eqnarray}
                    \gamma + \kappa & > & 0, \\
                    \frac{1}{4} \left( \gamma^{2} + 4 \gamma \kappa + \kappa^{2} \right) + 2 G_{-} G_{+} + \tilde{G}_{-} \tilde{G}_{+} & > & 0, \\
                    \tilde{G}_{-} \tilde{G}_{+} \kappa + \left( \frac{1}{4} \gamma \kappa +G_{-} G_{+} \right) \left( \gamma + \kappa \right) & > & 0, \\
                    \frac{\kappa^{2}}{16}  \left( 4 \tilde{G}_{-} \tilde{G}_{+} + \gamma^{2} \right) + G_{-}^{2} G_{+}^{2} + \frac{\gamma \kappa}{2} G_{-} G_{+} & > & 0.
                \end{eqnarray}
            \end{subequations}


    \section{Generation of Mechanical Squeezing}
        \label{sec:squeeze}
        \begin{figure}[ht]
            \centering
            \includegraphics[width=0.48\textwidth]{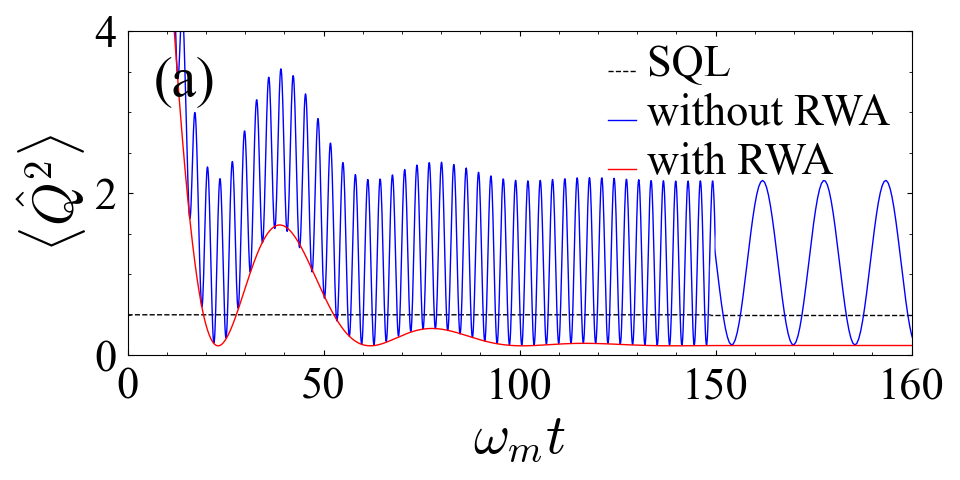}
            \includegraphics[width=0.48\textwidth]{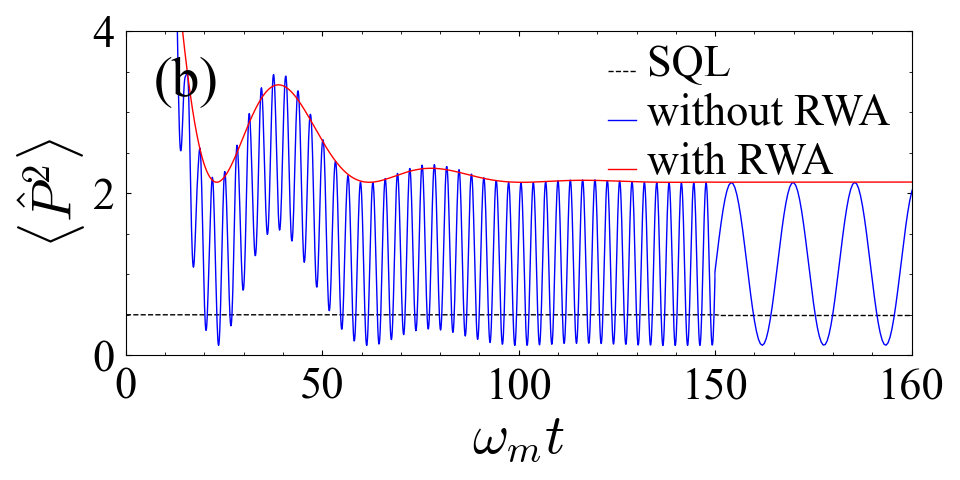}
            \caption{(Color online) Time-evolution of the variance in (a) position fluctuation ($Q$) and (b) momentum fluctuation ($P$) with RWA (solid red) and without RWA (solid blue).
            The dashed black line denotes the standard quantum limit.
            The parameters used are \cite{PhysRevA.101.053836} $\Delta = \omega_{m}$, $g = 10^{-4} \omega_{m}$, $\Omega_{a} = \Omega_{b} = 2 \omega_{m}$, $\kappa = 0.1 \omega_{m}$, $\gamma = 10^{-6} \omega_{m}$, $a_{0} = 2.0$, $a_{\pm 1} = 0.8$, $b_{0} = 100$, $b_{-1} = 25$, $b_{1} = 62.5$, $n_{a} = 0$ and $n_{b} = 10$.}
            \label{fig:squeeze_dyna}
        \end{figure}

        Fig. \ref{fig:squeeze_dyna} shows the time-evolution of the variances of the position and momentum quadrature fluctuations in the presence of drive amplitude modulation.
        Without RWA, steady-state squeezing of the quadrature occurs in the long-time limit, with a period determined by the modulation frequencies $\Omega_a$ and $\Omega_b$.
        This is in agreement with Eqs. \eqref{eqn:system_amp}.
        Under RWA, the variance in the position quadrature falls below the standard quantum limit.
        In both these cases, the maximum amount of squeezing achieved in the position quadrature is the same.

        \begin{figure*}
            \centering
            \includegraphics[width=0.184\textwidth]{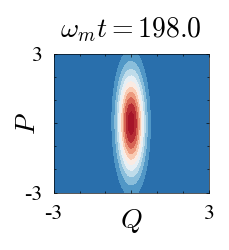}
            \includegraphics[width=0.184\textwidth]{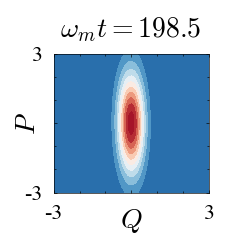}
            \includegraphics[width=0.184\textwidth]{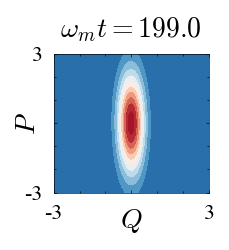}
            \includegraphics[width=0.184\textwidth]{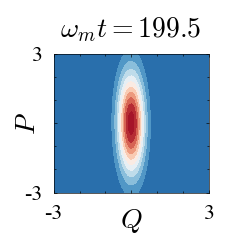}
            \includegraphics[width=0.24\textwidth]{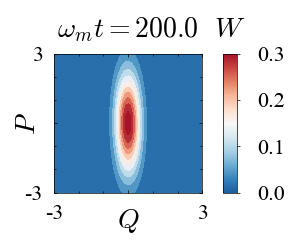}
            \includegraphics[width=0.184\textwidth]{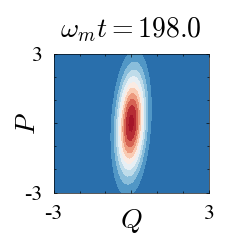}
            \includegraphics[width=0.184\textwidth]{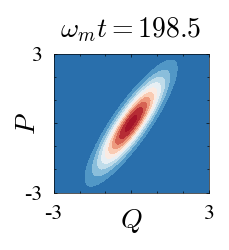}
            \includegraphics[width=0.184\textwidth]{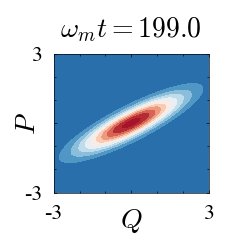}
            \includegraphics[width=0.184\textwidth]{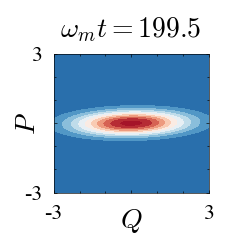}
            \includegraphics[width=0.24\textwidth]{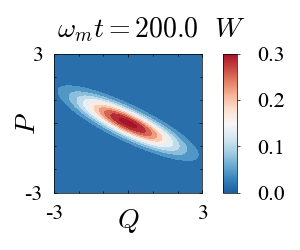}
            \caption{(Color online) Wigner function distribution of the mechanical mode at specific time intervals with (top panel) and without (bottom panel) RWA.
            System parameters are same as in Fig. \ref{fig:squeeze_dyna}.}
            \label{fig:squeeze_wigner}
        \end{figure*}
        
        Next, we seek to understand the effects of the nonresonant terms without RWA more succinctly by analyzing the Wigner function distribution of the mechanical quadratures in the phase space. 
        Given the Gaussian nature of the quantum noise, our linearized system can represented by a single mode Gaussian Wigner function in the steady-state as \cite{RevModPhys.84.621}
        \begin{eqnarray}
            \label{eqn:gauss_wig}
            W(\textbf{u}_{b}) = \frac{1}{2 \pi \sqrt{\textrm{det} \left[ \textbf{V}_{b} \right]}} \textrm{exp} \left[ {- \frac{\textbf{u}_{b}^{T}\textbf{V}_{b}^{-1} \textbf{u}_{b}}{2}} \right],
        \end{eqnarray}
        where $\textbf{u}_{b} = ( Q, P )^{T}$ is the vector of the mechanical fluctuations and $\textbf{V}_{b}$ is the covariance matrix for the mechanical mode.

        In Fig. \ref{fig:squeeze_wigner}, we plot the Wigner function distribution of the mechanical quadratures in the long-time limit i.e. $\omega_{m} t \to 1000$.
        We note that the direction of the squeezed quadrature rotates in the phase space due to the presence of the fast-rotating terms. 
        The period of this rotation $\tau$ is in accordance with the Floquet theorem mentioned in Sec. \ref{sec:system}. 
        Under RWA, the Wigner function does not display any rotation due to the absence of the fast-rotating terms.
        Also, the shape of the Wigner function in Fig. \ref{fig:squeeze_wigner} is preserved in both of these cases, indicating that the magnitude of squeezing is the same.

        \subsection{Squeezing as a Result of Competing Dynamics}
            \label{subsec:squeeze_comp}
            We now present an intuitive way of understanding the steady-state squeezing generated in our system.
            For equal values of $G_{0}$ and $G_{1}$, the cavity is coupled to the mechanical quadrature $X_{b}$.
            When $G_{0} \neq G_{1}$, the coupling between the cavity and the mechanical modes can be visualized using the Bogoliubov mode operator $\beta = \cosh{r} \tilde{b} + \sinh{r} \tilde{b}^{\dagger}$ where $\tanh{r} = G_{1} / G_{0}$ \cite{PhysRevA.88.063833}.
            The QLEs in Eqs. \eqref{eqn:system_rwa_qles} become,
            \begin{subequations}
                \begin{eqnarray}
                    \label{eqn:squeeze_comp_bog_a} 
                    \dot{\tilde{a}} & = & i \mathcal{G} \beta - \frac{\kappa}{2} \tilde{a} + \sqrt{\kappa} \tilde{a}_{in}, \\
                    \label{eqn:squeeze_comp_bog_b} 
                    \dot{\beta} & = & i \mathcal{G} \tilde{a} - \frac{\gamma}{2} \beta + \sqrt{\gamma} \beta_{in} \nonumber \\
                    && + i \left\{ \left( \frac{G_{0}^{2} + G_{1}^{2}}{\mathcal{G}^{2}} \right) \tilde{G}_{0} - \frac{2 G_{1} G_{0}}{\mathcal{G}^{2}} \tilde{G}_{1} \right\} \beta \nonumber \\ 
                    && + i \left\{ \left( \frac{G_{0}^{2} + G_{1}^{2}}{\mathcal{G}^{2}} \right) \tilde{G}_{1} - \frac{2 G_{1} G_{0}}{\mathcal{G}^{2}} \tilde{G}_{0} \right\} \beta^{\dagger},
                \end{eqnarray}
            \end{subequations}
            where $\mathcal{G} = \sqrt{G_{0}^{2} - G_{1}^{2}}$ is the effective coupling between the cavity mode and the Bogoliubov mode and $\beta_{in} = \cosh{r} \tilde{b}_{in} + \sinh{r} \tilde{b}_{in}^{\dagger}$ is the corresponding noise operator.
            \begin{figure}[ht]
                \centering
                \includegraphics[width=0.45\textwidth,height=4cm,keepaspectratio=False]{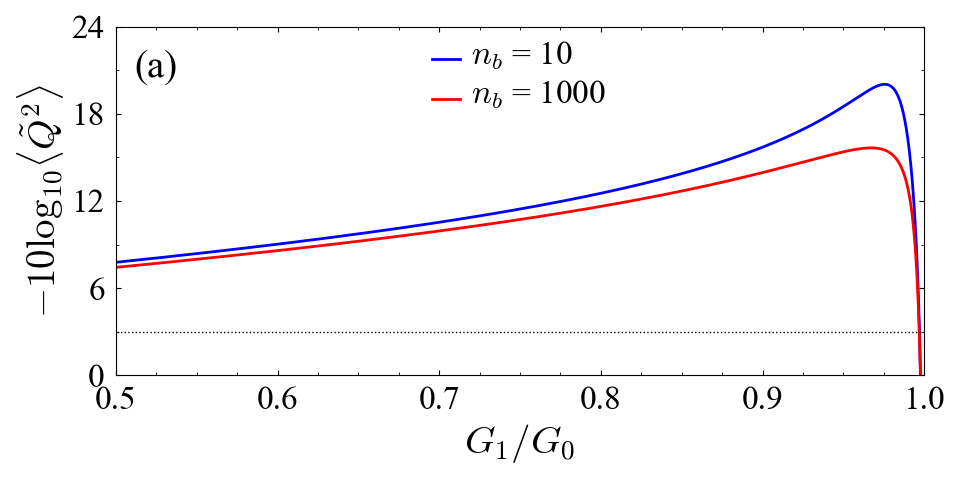}
                \includegraphics[width=0.45\textwidth,height=4cm,keepaspectratio=False]{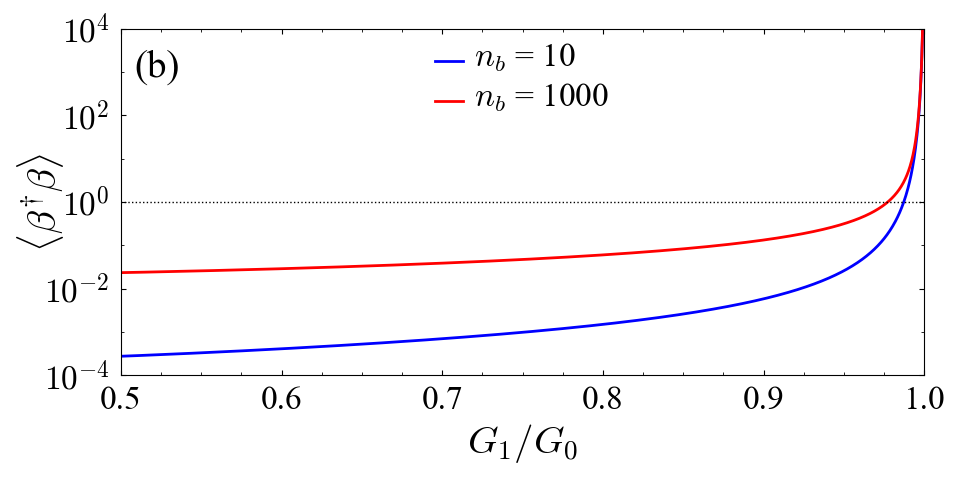}
                \caption{(Color online) Behaviour of (a) variance in the position quadrature $\langle \tilde{Q}^{2} \rangle$ and (b) occupancy of the Bogoliubov mode $\langle \beta^{\dagger} \beta \rangle$ with variation in the ratio of the sideband coupling strengths $G_{1} / G_{0}$ for $n_{b} = 10$ (solid blue) and $n_{b} = 1000$ (solid red).
                Here, we have used $(b_{-1} + b_{1}) \in [ 0, 600.0 ]$ with $b_{1} / b_{-1} = 2.5$. 
                Other parameters are same as in Fig. \ref{fig:squeeze_dyna}.
                The dotted black line in (a) denotes the SQL and the region under the dotted black line in (b) corresponds to cooling of the Bogoliubov mode ($\langle \beta^{\dagger} \beta \rangle < 1$).}
                \label{fig:squeeze_comp}
            \end{figure}
            
            In Fig. \ref{fig:squeeze_comp}(a), we observe a good amount of squeezing in the mechanical position for a wide range of the ratio $G_{1} / G_{0}$.
            However, as its value approaches unity, the occupancy of the Bogoliubov mode shoots up (refer Fig. \ref{fig:squeeze_comp}(b)), resulting in very low amount of coupling $\mathcal{G}$ between the modes.
            This in turn results in the mechanical squeezing to suddenly drop and thus, the corresponding variances exceed the SQL.
            
            To ensure stability, from Eqs. \eqref{eqn:system_rhc} we can reduce the constraint to $G_1< G_0$.
            Substituting the Bogoliubov operator in the linearized Hamiltonian, the interaction part of the resultant Hamiltonian takes the following form
            \begin{eqnarray}
                \mathcal{H} = - \mathcal{G} \left( \tilde{a}^{\dagger} \beta + \tilde{a} \beta^{\dagger} \right)
            \end{eqnarray}
            This is the well known beam-splitter Hamiltonian which describes the energy-exchange between the mechanical and the cavity mode. 
            It is widely applied to implement optomechanical sideband cooling of the mechanical mode \cite{PhysRevLett.99.093901, PhysRevLett.99.093902}. 
            Thus, the Bogoliubov mode $\beta$ undergoes ground-state cooling by interacting with the cavity mode $\tilde{a}$. 
            The squeezing parameter $r = \tanh^{-1} [ G_{1} / G_{0} ]$ increases with an increase in $G_{1}$ for a given $G_{0}$. 
            This causes the squeezing of the mechanical mode to be enhanced, which is clear from Fig. \ref{fig:squeeze_comp}(a). 
            The enhanced coupling rate between the optical field and the mechanical oscillator is now given by $\mathcal{G} = \sqrt{G_{0}^{2} - G_{1}^{2}}$. 
            With an increase in $G_{1}$, this value steadily decreases and finally becomes zero. 
            At this point the Bogoliubov cooling can no longer happen, which is shown by the sharp increase in its occupancy $\beta^{\dagger} \beta$ in Fig. \ref{fig:squeeze_comp}(b) as $G_{1} / G_{0} \to 1$.
            After this stage, the effects of thermal noise kick in, making it impossible to achieve high degree of squeezing. 
            This is shown by the abrupt fall in the value of $\langle \tilde{Q}^{2} \rangle$ almost competing with the rise in occupancy of the bogoliubov mode. 
            Thus, there is an optimum value of the sideband ratio $G_{1} / G_{0}$ for which the squeezing achieved in maximum.

        \subsection{Optimal ratio of the sideband strengths}
            \label{subsec:squeeze_opt}
            It can be seen from Fig. \ref{fig:squeeze_comp}(a) that for a certain value of the sideband ratio, the mechanical squeezing in the position quadrature is maximum.
            Also, it is well known that the degree of squeezing obtained is also a function of the dissipation of the cavity mode $\kappa$ \cite{PhysRevA.88.063833}.
            We now seek to optimize this ratio $G_{1} / G_{0}$ by plotting it as a function of the dissipation strength $\kappa$.
            
\begin{figure}[ht]
                \centering
                \includegraphics[width=0.45\textwidth,height=4cm,keepaspectratio=False]{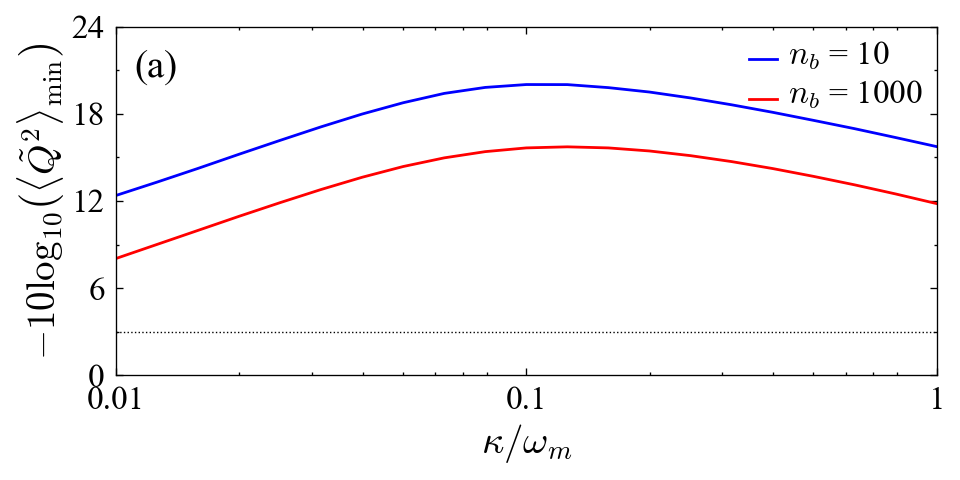}
                \includegraphics[width=0.45\textwidth,height=4cm,keepaspectratio=False]{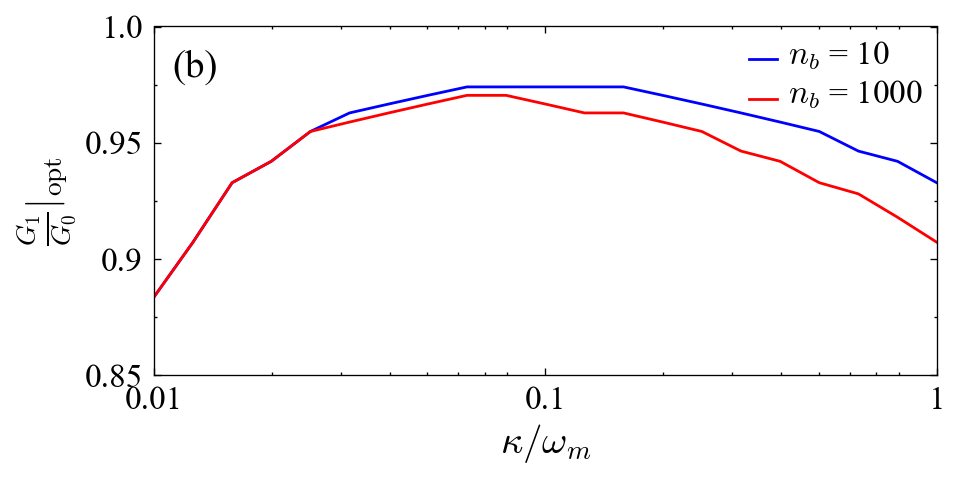}
                \caption{(Color online) (a) Maximum squeezing in the position quadrature and (b) optimal sideband ratios plotted with variation in the coupling strength of the sideband $G_{0}$ for $n_{b} = 10$ (solid blue) and $n_{b} = 1000$ (solid red).
                Other parameters are same as in Fig. \ref{fig:squeeze_dyna}.
                \label{fig:squeeze_opt}}
            \end{figure}

            In Fig. \ref{fig:squeeze_opt}(a), we show that the minimum quadrature variance $\langle \tilde{Q}^{2} \rangle_{\textrm{min}}$ for a specific value of the sideband ratio increases with an increase in $\kappa$ only up to a certain threshold. 
            This can be explained from the fact that the cooling of the Bogoliubov mode is enhanced, as the cavity dissipation $\kappa$ increases, for a specific sideband ratio.
            As $\kappa$ is increased further, the magnitude of squeezing decays.
            In Fig. \ref{fig:squeeze_opt}(b), we numerically optimize the sideband ratio $G_{1}/G_{0}$ by plotting its value corresponding to the highest degree of squeezing obtained in Fig. \ref{fig:squeeze_comp}(a) with respect to $\kappa$. 
            This behaviour is demonstrated by the tendency of $G_{1} / G_{0}$ to increase to a value close to 1, which corresponds to the maximized squeezing in Fig. \ref{fig:squeeze_opt}(a) and then gradually decrease for higher values of dissipation.
            
            One must note that, with an increase in the phonon number, the maximized $\langle \tilde{Q}^{2} \rangle$ decreases, due to the effect of thermal noise. 
            For our system, we show that for a hundred-fold increase in the thermal occupancy, the amount of squeezing obtained remains significantly above the SQL.
            This claim is further bolstered in Fig. \ref{fig:robust}, where we can see that the robustness of this scheme is much better than the results obtained so far \cite{PhysRevA.101.053836}.

        
    \section{Analytical and Numerical Solutions}
        \label{sec:soln}
        \subsection{Analytical Solution}
            \label{subsec:soln_analytic}
            In the weak optomechanical coupling limit (i.e. $\kappa \gg \mathcal{G}$), the cavity photons decay much faster than the optomechanical interaction.
            The cavity field therefore follows the mechanical motion adiabatically.
            Using this approximation to eliminate the cavity mode, i.e., setting $\dot{\tilde{a}} = 0$ \cite{PhysRevA.93.043844}, we solve for the variance of the position quadrature of the mechanical mode at the steady state of the cavity mode,
            \begin{subequations}
                \begin{eqnarray}
                    \label{eqn:soln_analytic_a} 
                    \tilde{a} & = & \frac{2 i \mathcal{G}}{\kappa} \beta + \frac{2}{\sqrt{\kappa}} \tilde{a}_{in}, \\
                    \label{eqn:soln_analytic_b} 
                    \dot{\beta} & = & - h \beta + \frac{2 i \mathcal{G}}{\sqrt{\kappa}} \tilde{a}_{in} + \sqrt{\gamma} \beta_{in} \nonumber \\
                    && + i \left\{ \left( \frac{G_{0}^{2} + G_{1}^{2}}{\mathcal{G}^{2}} \right) \tilde{G}_{0} - \frac{2 G_{1} G_{0}}{\mathcal{G}^{2}} \tilde{G}_{1} \right\} \beta \nonumber \\ 
                    && + i \left\{ \left( \frac{G_{0}^{2} + G_{1}^{2}}{\mathcal{G}^{2}} \right) \tilde{G}_{1} - \frac{2 G_{1} G_{0}}{\mathcal{G}^{2}} \tilde{G}_{0} \right\} \beta^{\dagger},
                \end{eqnarray}
            \end{subequations}
            where $h = \frac{2 \mathcal{G}^{2}}{\kappa} + \frac{\gamma}{2}$.
            From here, we obtain the steady-state solution of the variance in the position quadrature as (refer Appendix \ref{app:soln_analytic} for detailed derivation)
            \begin{eqnarray}
                \label{eqn:soln_analytic_pos} 
                \langle \tilde{Q}^{2} \rangle_{s} & = & \frac{h e^{-2 r}}{2 \left( \tilde{G}^{2}_{-} - h^{2} \right)} \Bigg\{ \gamma \left( n_{b} + \frac{1}{2} \right) \left( \frac{\tilde{G}_{-} G_{+}}{G_{-} h} e^{-2 r} - e^{2 r} \right) \nonumber \\
                && - \frac{4 \mathcal{G}^{2}}{\kappa} \left( n_{a} + \frac{1}{2} \right) \left( 1 + \frac{\tilde{G}_{-} G_{+}}{G_{-} h} \right) \Bigg\},
            \end{eqnarray}


        \subsection{Numerical Solution}
            \label{subsec:soln_numeric}
            To verify the accuracy of the analytical form derived under the adiabatic approximation, we explore the numerical solution for the steady-state position variance $\langle \tilde{Q}^{2} \rangle$.
            In the Fourier domain, Eq. \eqref{eqn:system_rwa_u} can be written as
            \begin{eqnarray}
                \label{eqn:soln_numeric_u}
                \tilde{u} ( \omega ) = ( i \omega \mathbb{I} + \tilde{M} )^{-1} \tilde{N} ( \omega ).
            \end{eqnarray}

            This gives us the expression for the position fluctuation of the mechanical mode,
            \begin{eqnarray} 
            \label{eqn:num_pos_fluc}
                \tilde{Q} & = & A ( \omega ) \tilde{X}_{in} + B ( \omega ) \tilde{Y}_{in} + E ( \omega ) \tilde{Q}_{in} + F ( \omega ) \tilde{P}_{in}.
            \end{eqnarray}

            The first two terms in the above expression arise from the vacuum noise in the cavity mode, and the last two terms arise from the thermal noise in the mechanical mode.
            We can then obtain the steady-state variance of the position quadrature $\langle \tilde{Q}^{2} \rangle_{s}$ by integrating its fluctuation spectra $S_{\tilde{Q}} ( \omega )$ within the whole spectral domain (refer Appendix \ref{app:soln_numeric} for detailed derivation),
            \begin{eqnarray}
            \label{eqn:num_sol_int}
                \langle \tilde{Q}^{2} \rangle_{s} = \frac{1}{2 \pi} \int_{- \infty}^{\infty} d \omega S_{\tilde{Q}} ( \omega ).
            \end{eqnarray}
            
\begin{figure}[ht]
                \centering
                \includegraphics[width=0.48\textwidth]{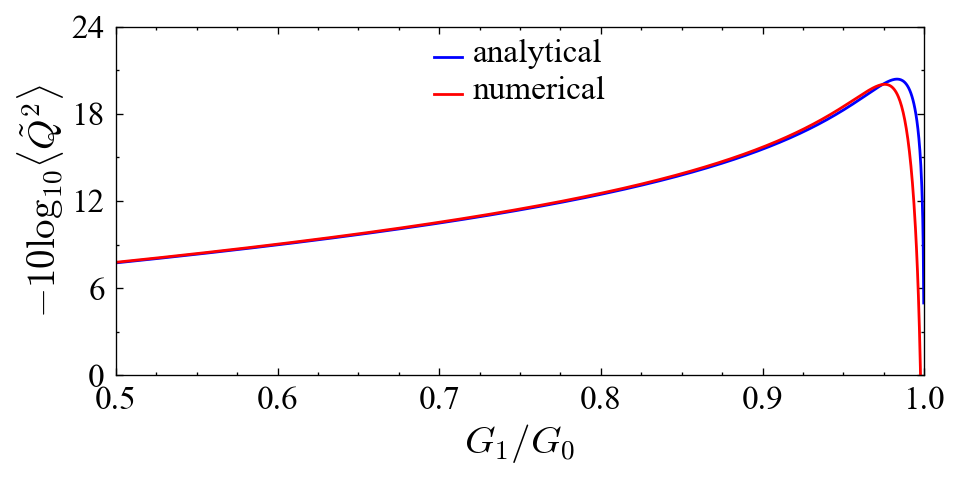}
                \caption{(Color online) Behaviour of squeezing in the position quadrature $\langle \tilde{Q}^{2} \rangle$ with variation in the sideband ratio $G_{1} / G_{0}$ for $n_{b} = 10$ obtained using the approximate solution Eq. \eqref{eqn:soln_analytic_pos} (solid blue) and the numerical solution of $\langle \tilde{Q}^{2} \rangle_{s} = \frac{1}{2 \pi} \int_{- \infty}^{\infty} d \omega S_{\tilde{Q}} ( \omega )$ (solid red).
                Other parameters are same as in Fig. \ref{fig:squeeze_comp}.}
                \label{fig:soln}
            \end{figure}

            In Fig. \ref{fig:soln}, we plot the numerically obtained steady-state position variance with respect to the sideband ratio alongside the analytical solution obtained in Eq. \eqref{eqn:soln_analytic_pos} for thermal phonon number $n_{b} = 10$. 
            It is evident from the figure that the behaviour of the analytical (from Eq. \eqref{eqn:soln_analytic_pos}) as well as the numerical solution (from Eq. \eqref{eqn:num_sol_int}) match very well with the one shown in Fig. \ref{fig:squeeze_comp}(a).
            
    \section{Robustness of Squeezing and Entanglement}
        \label{sec:robust}
\begin{figure}[ht]
            \centering
            \includegraphics[width=0.48\textwidth]{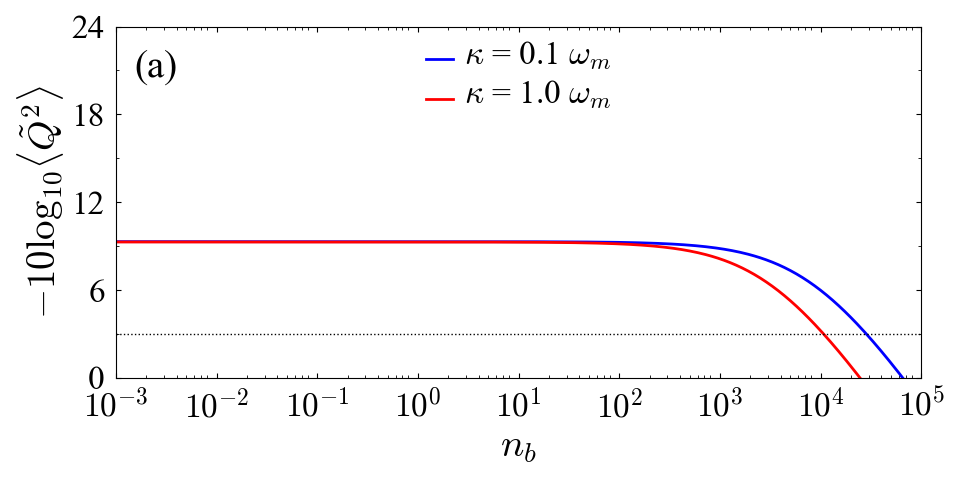}
            \includegraphics[width=0.48\textwidth]{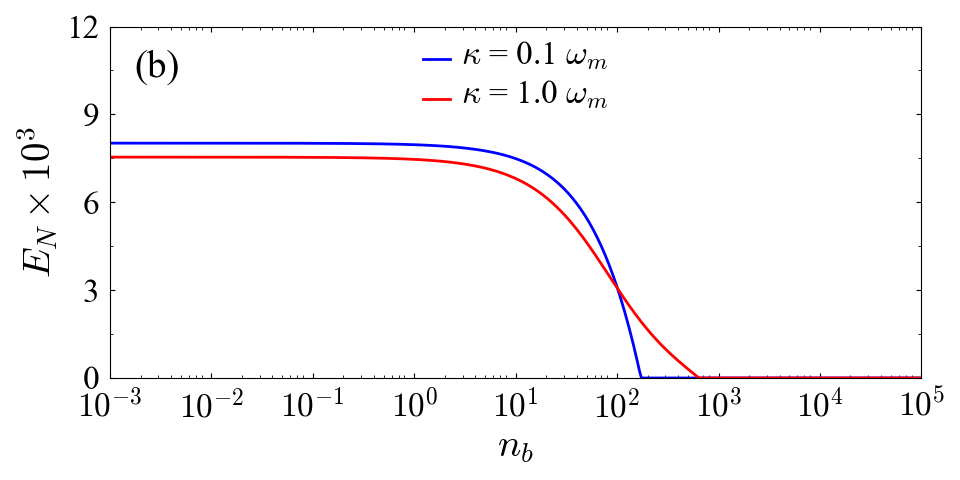}
            \includegraphics[width=0.48\textwidth]{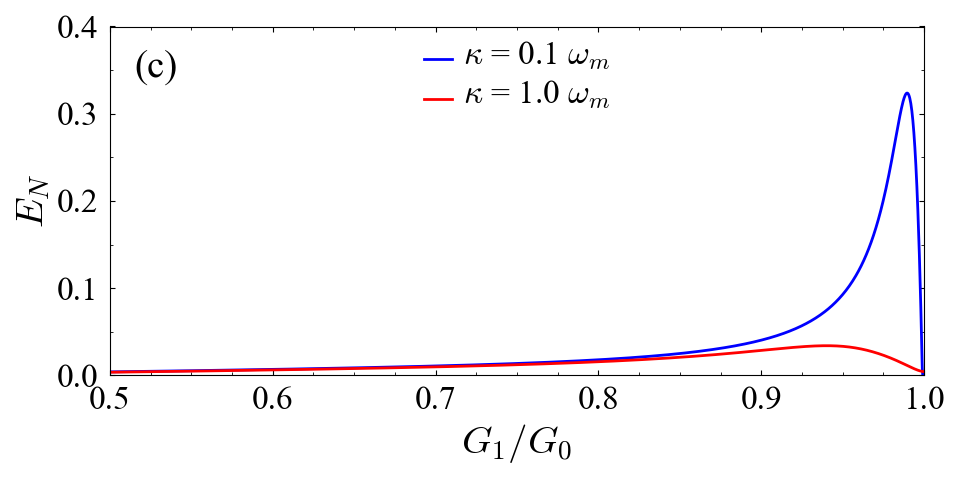}
            \caption{(Color online) (a) Steady state variance in position quadrature $\langle \tilde{Q}^{2} \rangle_{s}$ and (b) entanglement between the optical and mechanical modes $E_{N}$ with varying occupancy of the mechanical mode $n_{b}$ for $\kappa \in \{ 0.1, 1.0 \} \omega_{m}$.
            (c) Nature of entanglement $E_{N}$ with change in sideband coupling strength $G_{1} / G_{0}$ for different values of $\kappa$.
            Other parameters are same as in Fig. \ref{fig:squeeze_dyna}.
            The dotted black line in (a) denotes the SQL.}
            \label{fig:robust}
        \end{figure}

        The thermal occupancy and the accompanying position fluctuations of the mechanical mode is substantially higher than that of the cavity photons, even at a relatively low temperature.
        To this end, we check the robustness of the mechanical squeezing by plotting $\langle \tilde{Q}^{2} \rangle$ obtained using Eq. \eqref{eqn:system_rwa_M} as a function of the thermal phonon occupation number $n_{b}$ in Fig. \ref{fig:robust}(a).
        A strong mechanical squeezing is achieved for a low occupancy of the thermal bath and is robust upto a few thousands of thermal occupancies.

        Alongside squeezing, it is important to present the entanglement between the optical and mechanical modes of our system.
        To quantify the degree of entanglement, we use the standard logarithmic negativity measure $E_{N}$ \cite{PhysRevLett.98.030405} (refer Appendix \ref{app:entan} for its implementation).
        In Fig. \ref{fig:robust}(b), we plot this measure for different thermal occupancies and cavity linewidths.
        Similar to the behaviour of squeezing, we see that the entanglement between the modes is robust upto a few tens of thermal photons.
        However, the value of observed entanglement is very small for smaller values of $G_{1} / G_{0}$ and greatly enhances when the ratio is close to unity.

    \section{Conclusion}
        \label{sec:conc}
        In the present work, we have demonstrated the combined effect of pump modulation and cavity dissipation for generation of mechanical squeezed states in a optomechanical membrane-in-the-middle system. 
        By introducing a periodic modulation in the amplitude of the external laser field, we observed that the coupling between the mechanical and the optical mode takes a form which contributes to the ground-state cooling of the Bogoliubov mode.
        We investigated the time evolution of the squeezed position quadrature of the mechanical mode, both with and without considering the RWA. 
        We then concluded from its Wigner distribution that we can safely ignore the fast-oscillating terms using the RWA given that the degree of quadrature squeezing remains the same in both cases, besides a phase space rotation for the latter case.
        Then we analyzed how the magnitude of squeezing depends on the ratio of the coupling sideband and numerically optimized this ratio to obtain maximum squeezing for a value of $G_{1} / G_{0} \to 1$.
        Next, we found the analytical and numerical solution to obtain the steady-state quadrature variance, which matched quite well with our previously obtained results.
        A key outcome of our study is the robustness of the degree of mechanical squeezing and optomechanical entanglement which stays well above the SQL even for a high occupancy of the thermal bath.
        It is worth noting that this scheme successfully avoids parametric instability which might result from two driving tones by using a single-tone laser driving. 
        The scheme presented here may thus be very useful for designing instruments that carry out ultraprecise measurements based on mechanical squeezing (e.g., the gravitational-wave detectors) or even detecting quantum effects at a macroscopic scale.
        

    \section*{Acknowledgement}
        SK would like to acknowledge MHRD, Government of India for providing financial support for his research via the PMRF scheme.


    \appendix

    \section{Analytical Solution of Position Squeezing}
        \label{app:soln_analytic}
        The fluctuation in the quadrature operators in the Bogoliubov Mode can be calculated from Eq. \eqref{eqn:soln_analytic_b} as  \\
        \begin{subequations}
            \begin{eqnarray}
                \label{eqn:quad_bogo_a}
                \dot{X}_{\beta} & = & -h X_{\beta}-\frac{2}{\sqrt{\kappa}}\mathcal{G}\tilde{Y}^{in}(t)+\sqrt{\gamma}X_{\beta}^{in} \nonumber \\ && -\Big(\frac{G_{+}}{G_{-}}\Big )\tilde{G}_{-} Y_{\beta}, \\
                \label{eqn:quad_bogo_b}
                \dot{Y}_{\beta} & = & -h Y_{\beta}+\frac{2}{\sqrt{\kappa}}\mathcal{G}\tilde{X}^{in}(t)+\sqrt{\gamma}Y_{\beta}^{in} \nonumber \\ &&  +\Big(\frac{G_{-}}{G_{+}}\Big )\tilde{G}_{+} X_{\beta}.
            \end{eqnarray}
        \end{subequations}
        Here, we can assign the terms
        $-\frac{2}{\sqrt{\kappa}}\mathcal{G}\tilde{Y}^{in}(t)$, $\sqrt{\gamma}X_{\beta}^{in}$ and $\sqrt{\gamma}Y_{\beta}^{in}$ to $\mathcal{F}_{1}(t), \mathcal{F}_{2}(t) \text{ and }\mathcal{F}_{3}(t)$ respectively, where $\mathcal{F}_i$ are the effective quantum langevin forces, whose correlations are given as
        \begin{subequations}
            \begin{eqnarray}
                \label{eqn:langev_corr_a}
                \langle \mathcal{F}_{1}(t)\mathcal{F}_1(t')\rangle & = & (-\frac{2\mathcal{G}}{\sqrt{\kappa}})^2(n_{a}+\frac{1}{2})\delta(t-t'), \\
                \label{eqn:langev_corr_b}
                \langle \mathcal{F}_{2}(t)\mathcal{F}_2(t')\rangle & = & \gamma e^{2r}(n_{b}+\frac{1}{2})\delta(t-t'),
                \\
                \label{eqn:langev_corr_c}
                \langle \mathcal{F}_{3}(t)\mathcal{F}_{3}(t')\rangle & = & -\gamma e^{-2r}(n_{b}+\frac{1}{2})\delta(t-t').
            \end{eqnarray}
        \end{subequations}
        At steady-state, taking $\langle \dot{X}_{\beta}^2\rangle=\langle \dot{Y}_{\beta}^2\rangle=0$,
        \begin{subequations}
            \begin{eqnarray}
                \label{eqn:adb_quad_bogo_a}
                \langle  X_{\beta}^2\rangle_{s} & = & \frac{2\mathcal{G}^2}{\kappa h}(n_{a}+\frac{1}{2})+\frac{\gamma}{2h} e^{2r}(n_{b}+\frac{1}{2}) \nonumber \\ && -\Big(\frac{G_{+}}{G_{-}h}\Big )\tilde{G}_{-}\langle Y_{\beta}^2\rangle_{s}, \\
                \label{eqn:adb_quad_bogo_b}
                \langle  Y_{\beta}^2\rangle_{s} & = & \frac{2\mathcal{G}^2}{\kappa h}(n_{a}+\frac{1}{2})-\frac{\gamma}{2h} e^{-2r}(n_{b}+\frac{1}{2}) \nonumber \\ && +\Big(\frac{G_{-}}{G_{+}h}\Big )\tilde{G}_{+}\langle X_{\beta}^2\rangle_{s}.
            \end{eqnarray}
        \end{subequations}
        The coupled equations in Eqs. \eqref{eqn:adb_quad_bogo_a} and \eqref{eqn:adb_quad_bogo_b} can be solved to obtain $\langle  X_{\beta}^2\rangle$. The analytical solution for the variance in position quadrature at steady-state in Eq.  \eqref{eqn:soln_analytic_pos} is obtained from $\langle\tilde{Q}^2\rangle_s =e^{-2r}\langle  X_{\beta}^2\rangle_{s}$.
    \section{Numerical Solution of Position Squeezing}
        \label{app:soln_numeric}
        The coefficients in Eq. \eqref{eqn:num_pos_fluc} are,
        \begin{widetext}
            \begin{subequations}
            \begin{eqnarray}
                \label{eqn:num_coeff}
                A(\omega) & = & -\frac{8 G_+ \sqrt{\kappa } \tilde{G}_- (\kappa -2 i \omega )}{(\kappa -2 i \omega )^2 \left(-4 \tilde{G}_- \tilde{G}_++\left(2 \omega +i \gamma _m\right){}^2\right)+8 G_- G_+ (2 \omega +i \kappa ) \left(2 \omega +i \gamma _m\right)-16 G_-^2 G_+^2},\\
                B(\omega) & = & \frac{4 G_- \sqrt{\kappa } \left(-4 G_- G_++(2 \omega +i \kappa ) \left(2 \omega +i \gamma _m\right)\right)}{(\kappa -2 i \omega )^2 \left(-4 \tilde{G}_- \tilde{G}_++\left(2 \omega +i \gamma _m\right){}^2\right)+8 G_- G_+ (2 \omega +i \kappa ) \left(2 \omega +i \gamma _m\right)-16 G_-^2 G_+^2},\\
                C(\omega) & = & \frac{2 (\kappa -2 i \omega ) \sqrt{\gamma _m} \left(4 G_- G_++(\kappa -2 i \omega ) \left(\gamma _m-2 i \omega \right)\right)}{(\kappa -2 i \omega )^2 \left(-4 \tilde{G}_- \tilde{G}_++\left(2 \omega +i \gamma _m\right){}^2\right)+8 G_- G_+ (2 \omega +i \kappa ) \left(2 \omega +i \gamma _m\right)-16 G_-^2 G_+^2},\\
                D(\omega) & = & -\frac{4 \tilde{G}_- (\kappa -2 i \omega )^2 \sqrt{\gamma _m}}{(\kappa -2 i \omega )^2 \left(-4 \tilde{G}_- \tilde{G}_++\left(2 \omega +i \gamma _m\right){}^2\right)+8 G_- G_+ (2 \omega +i \kappa ) \left(2 \omega +i \gamma _m\right)-16 G_-^2 G_+^2}.
            \end{eqnarray}
            \end{subequations}
        \end{widetext}
        The fluctuation spectrum of the position quadrature of the mechanical mode $ \tilde{Q}$ can be written as
        \begin{eqnarray}
            S_{\tilde{Q}}(\omega)=\frac{\langle \tilde{Q}(\Omega)\tilde{Q}(\omega)\rangle +\langle \tilde{Q}(\omega)\tilde{Q}(\Omega)\rangle}{4\pi\delta(\omega+\Omega)}.
        \end{eqnarray}
        Using the correlations between the noise operators $\tilde{\mathcal{O}}_{a(b)}^{in}(\omega)$, the fluctuation spectrum takes the form,
        \begin{eqnarray}
            S_{\tilde{Q}}(\omega) & = &  [A(-\omega) A(\omega)+B(-\omega) B(\omega)](n_a+\frac{1}{2})\nonumber \\ && +[C(-\omega) C(\omega)+D(-\omega) D(\omega)](n_b+\frac{1}{2}).
        \end{eqnarray}
    \section{Measure of Entanglement}
        \label{app:entan}
        The bipartite entanglement between the optical and mechanical mode is obtained numerically by writing the correlation matrix in the standard form
        \begin{eqnarray}
            \textbf{V} (t) = \begin{pmatrix}
                \mathcal{A} (t) & \mathcal{C} (t) \\
                \mathcal{C}^{T} (t) & \mathcal{B} (t)
            \end{pmatrix},
        \end{eqnarray}
        where $\mathcal{A} (t)$, $\mathcal{B} (t)$ and $\mathcal{C} (t)$ are $2 \times 2$ matrices.
        The degree of entanglement is then quantified by the logarithmic negativity metric as \cite{PhysRevLett.98.030405}
        \begin{eqnarray}
            E_{N} (t) = \max{\left[ 0, - \ln {\left( 2 \nu^{-} (t) \right)} \right]},
        \end{eqnarray}
        where $\nu^{-} (t) = 2^{-1 / 2} \{ \Sigma (t) - \sqrt{\Sigma^{2} (t) - 4 \det{[ V_{bc} (t) ]}} \}^{1 / 2}$ with $\Sigma (t) \equiv \det{[ \mathcal{A} (t) ]} + \det{[ \mathcal{B} (t) ]} - 2 \det{[ \mathcal{C} (t) ]}$.

    \bibliography{references}

\end{document}